\documentclass[aps,prl,groupedaddress]{revtex4}

\usepackage{keyval} \usepackage{dcolumn}

\usepackage[]{amsmath}

\def\epl{Europhys.\ Lett. }

\def\prl{Phys.\ Rev.\ Lett.}
\def\prb{Phys.\ Rev.\ B }
\def\pr{Phys.\ Rev. }

\def\rmp{Rev.\ Mod.\ Phys. }

\usepackage[psamsfonts]{amsfonts}
\usepackage[psamsfonts]{amssymb}
	
\usepackage{graphicx}

\usepackage{bm}

\usepackage[usenames,dvipsnames]{color}

\usepackage[normalem]{ulem}

\newcommand{\threej}[6]{\ensuremath{\left(\begin{array}{ccc}#1 &#2 &#3 \\ #4 &#5 &#6 \end{array}\right)}}
\newcommand{\sixj}[6]{\ensuremath{\left\{\begin{array}{ccc}#1 &#2 &#3 \\ #4 &#5 &#6 \end{array}\right\} }}

\newcommand{\comment}[1]{\textcolor{red}{#1}}
\renewcommand{\comment}[1]{\relax}
\newcommand{\tobedeleted}[1]{\textcolor{green}{\sout{#1}}}
\renewcommand{\tobedeleted}[1]{\relax}

\begin{document}

\title{Multipole decomposition of LDA+$U$ energy and its application to actinides compounds}

\author{Fredrik Bultmark}
\affiliation{Department of Physics and Materials Science, Uppsala University, Box 530, SE-75121 Uppsala, Sweden}
\author{Francesco Cricchio}
\affiliation{Department of Physics and Materials Science, Uppsala University, Box 530, SE-75121 Uppsala, Sweden}  
\author{Oscar Gr\aa n\"as}
\affiliation{Department of Physics and Materials Science, Uppsala University, Box 530, SE-75121 Uppsala, Sweden} 
\author{Lars Nordstr\"om}
\affiliation{Department of Physics and Materials Science, Uppsala University, Box 530, SE-75121 Uppsala, Sweden}

\date{\today}

\begin{abstract}
A general reformulation of the exchange energy of $5f$-shell  is applied in the analysis of the magnetic structure of various actinides compounds in the framework of LDA+$U$ method.  The calculations are performed in a convenient scheme with essentially only one free parameter, the screening length. The results are analyzed in terms of different polarization channels, due to different multipoles.
Generally it is found that the spin-orbital polarization is dominating. This can be viewed as a strong enhancement of the spin-orbit coupling in these systems. This leads to a drastic decrease in spin polarization, in accordance with experiments. The calculations are able to correctly differentiate magnetic and non-magnetic Pu system. Finally, in all magnetic systems a new multipolar order is observed, whose polarization energy is often larger in magnitude than one of spin polarization.
 \end{abstract}

\maketitle

\section{Introduction}
The magnetism of actinide systems shows a very rich variety of magnetic properties~\cite{HB}. There are variations from itinerant magnetic systems to systems showing characteristics of localized magnetism.
In the border between these extremes one have the so-called heavy fermions, which show many peculiar and anomalous properties, one of which is the co-existence of superconductivity and magnetism~\cite{Heavy-fermions}.
One aspect that makes the magnetism of the actinides unique is the presence of strong spin-orbit coupling (SOC) together with strong exchange interactions for the $5f$ electrons, which are the ones responsible for the magnetism.

From a theoretical point of view, a standard density functional approach, either in the local density approximation (LDA) or generalized gradient approximation (GGA), describes quite well the equilibrium properties of at least metallic systems. However, these functionals are known to underestimate the orbital moments which are induced by the relatively strong SOC~\cite{LDA-fails-I,LDA-fails-II,Brooks}.
This can be remedied by allowing for the so-called orbital polarization~\cite{Brooks}, responsible for Hund's second rule in atomic physics, either through adding an appropriate orbital depending term to the Hamiltonian or by adopting the so-called LDA+$U$ approach~\cite{Solovyev-I,Anisimov,LAZ}. In the latter method a screened Hartree-Fock (HF) interaction is included among the $5f$ states only.

There is a drastic difference between the itinerant magnetism of a $3d$ shell and that of the $5f$ shell. In the former the orbital degrees of freedom are quenched due to the process of hopping between different atoms, while in the latter case the stronger SOC un-quenches them again. Magnetic ordering is relatively abundant among actinide systems due to the strong exchange interactions, but generally the spin moments are strongly reduced compared to a fully spin polarized value, which sometimes is ascribed to crystal field effects and other times to hybridization.

This paper will focus on the role of the local screened exchange interactions and it will aim to convincingly 
argue that these interactions, together with an appreciable SOC interaction, are responsible for the reduced spin polarizations as well as for a large orbital moment. This is analyzed in terms of tensor moment polarizations of the $5f$ shell.
Firstly we describe the method we employ, the LDA+$U$ method in its most general form with a minimum of free parameters; one. This is accomplished by using screened Slater parameters together with an interpolating optimal double-counting (DC) term~\cite{Petukhov}. This last degree of freedom can be chosen to be, for instance, the lowest Slater integral $U$, which is used as a varying parameter. This general  LDA+$U$ scheme has been implemented within the state-of-the-art full-potential augmented plane-wave plus local orbitals (FP-APW+$lo$) method~\cite{APW+lo-I,APW+lo-II} in the Elk code~\cite{Elk}. This linearized augmented plane-wave method is one of the most accurate scheme to treat the complicate behavior of $5f$ electrons in actinides~\cite{Laskowski,Shick_UGe2}.
Secondly, we present an analysis method for the resulting ground state. This analysis is based on an exact multipole decomposition of the density matrix as well as of the HF exchange energy. We have tested the generality of the method by considering actinides systems with different degree of localization of their $5f$ electrons; the itinerant features in U compounds, the more localized behavior in Pu compounds and finally the intermediate one in Np based materials.
The calculations show reasonable comparison with experiments as well to other beyond-LDA calculations. By applying our tensor moment decomposition, we observe several clear trends regarding the favoured polarizations in these actinide systems. For instance it is clear that the Hund's rules break down in the sense that other channels are more significant than the spin polarization.

 \section{Method}
\subsection{A General Form of LDA+$U$ Method}
In the most general version of LDA+$U$ \cite{Solovyev-I,LAZ,Shick} a HF correction to the energy enters with the form
\begin{equation}
E_\mathrm{HF}=\frac{1}{2}\sum_{abcd} \rho_{ac} \left[\left<ab|g|cd\right> - \left<ab|g|dc\right>\right] \rho_{bd} \ , 
\label{eq:HF}
\end{equation}
where $\rho_{ab}$ is one element of the density matrix with dimension $D=2(2\ell+1)$, which acts as an occupation matrix and
$a$, $b$, $c$ and $d$ are single electron states.%
%
The interaction term is of the form
\begin{equation}
\left<ab|g|cd\right> = \int \psi_{a}^\dagger(1)\psi_{b}^\dagger(2) g(r_{12}) \psi_{c}(1)\psi_{d}(2) \mathrm{d(1) d(2)} \ ,\label{interaction}
\end{equation}
with one-electron states $a$ with wavefunction $\psi_{a}(1)=R_{\ell}(r_{1})Y_{\ell m_{a}}(\Omega_{1})\chi_{s_{a}}(1)$, where the relevant quantum numbers, $m$ and $s$, are the magnetic quantum number and spin component, respectively. The interaction can be expanded in a series
\begin{equation}
g(r_{12})=\sum_{k=0}^\infty g_{k}(r_{1},r_{2}) P_{k}(\cos \theta_{12})\ ,
\end{equation}
where the Legendre function $P_{k}$ in turn can be expanded by the use of the addition theorem for spherical harmonics
\begin{equation}
P_{k}(\cos \theta_{12})=\frac{4\pi}{2k+1}\sum_{q=-k}^{k} Y_{kq}^*(\Omega_{1})Y_{kq}(\Omega_{2}) \ .
\end{equation}
\subsection{Calculation of Slater Integrals through a Yukawa Potential}
The radial part of the interaction is then contained in the Slater integrals
\begin{equation}
\label{eq:Fks}
F^{(k)}=\int dr_{1} r_{1}^2 R_{\ell}^2(r_{1}) g_{k}(r_{1},r_{2}) R_{\ell}^2(r_{2})  r_{2}^2 dr_{2}\ .
\end{equation}
For a $f$ shell there are four independent parameters, $F^{(0)}=U$, $F^{(2)}$, $F^{(4)}$ and $F^{(6)}$. It is very unpractical to stay with these four parameters.
A common practice within  LDA+$U$ or HF calculations~\cite{LAZ,Shick,Shishidou} is to have the screened Slater parameters determined by the choice of two linear combinations of parameters, $U$ and $J$, and by fixing two ratios, $A_1={F^{(4)}}/{F^{(2)}}$ and $A_2={F^{(6)}}/{F^{(2)}}$.
In the present work we will instead follow the ideas of Norman~\cite{Norman} and calculate the Slater parameters directly from a screened Coulomb interaction  in the form of a Yukawa potential  $g(r_{1},r_{2})=
{e^{-\lambda r_{12}}}/{r_{12}}$. Then
\begin{equation}
g_{k}(r_{1},r_{2},\lambda)=-(2k+1)\lambda j_k(i\lambda r_{<})h_k^{(1)}(i\lambda r_{>})\ ,
\label{eq:Yukawa}
\end{equation}
where $j_k$ is a spherical Bessel function, $h_k^{(1)}$ is spherical Hankel function of the first kind and $r_{<}$ and $r_{>}$ are, respectively, the smaller and the larger radius entering in the double integral in Eq.~(\ref{eq:Fks}). This type of approach  has two advantages; it determines the ratio between the different Slater parameters in a more realistic way than by choosing $U$ and $J$ individually, and there is only one independent parameter, the screening length $\lambda$.

\begin{figure}[htbp]
\begin{center}
\includegraphics[width=0.45\columnwidth]{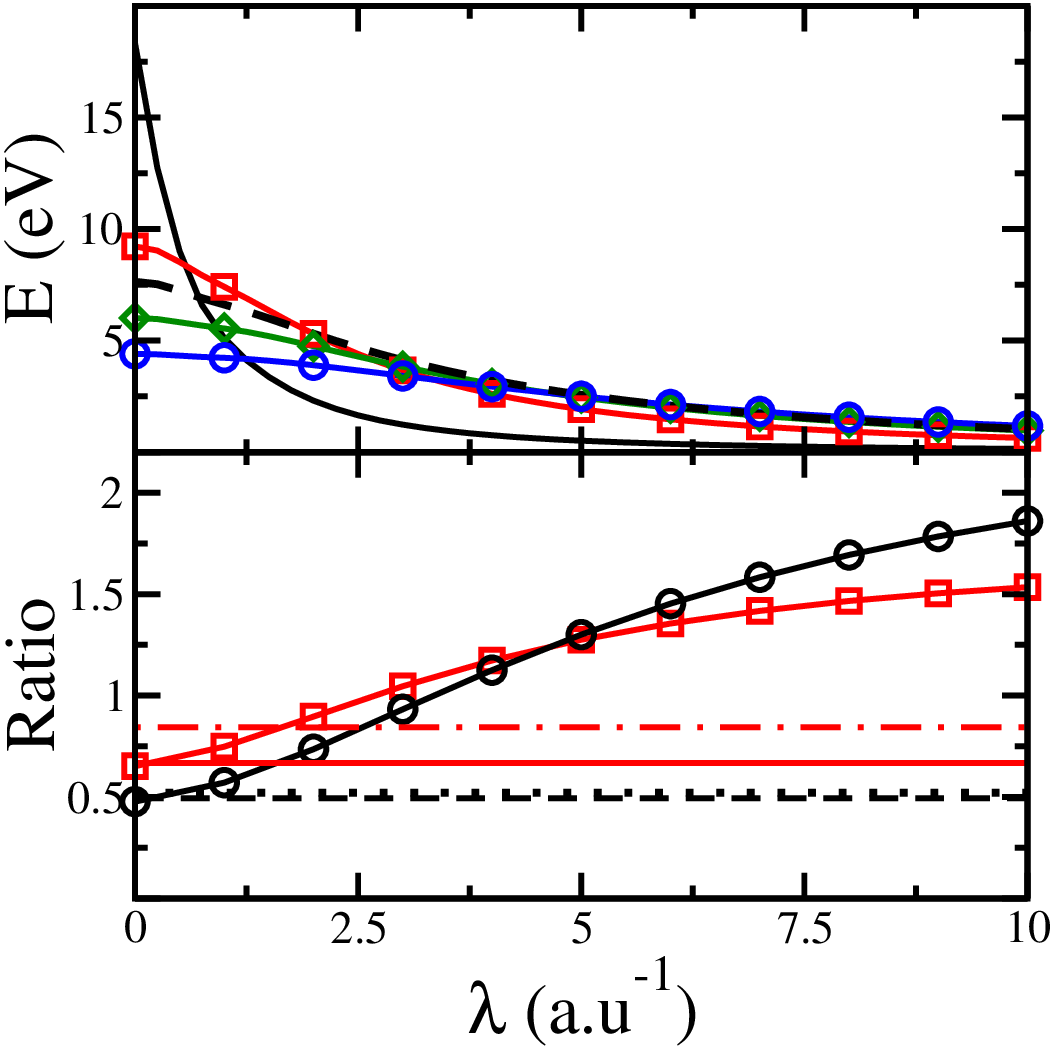}
\caption{Upper panel: Slater parameters $F^{(0)}$ (full black line), $F^{(2)}$ (red line with squares),  $F^{(4)}$ (green line with diamonds), $F^{(6)}$ (blue line with circles) and parameter $J$ times 10 (dashed black line) as function of screening length $\lambda$ of the Yukawa potential in US. Lower panel: comparison of Slater parameters ratios $A_1=F^{(4)}/F^{(2)}$ (red lines) and  $A_2=F^{(6)}/F^{(2)}$ (black lines) calculated by a screened Yukawa potential (full line with squares for $A_1$, full line with circles for $A_2$) with the fixed  ones of Ref.~\cite{LAZ,Judd} (full line for $A_1$, dashed line for $A_2$) and of Ref.~\cite{Shishidou,Norman} (dashed-dotted line for $A_1$, dotted line for $A_2$). The radial wave-function used for the calculation of Slater integrals in Eq.~(\ref{eq:Fks}) corresponds to the energy at the center of $5f$ band and the MT radius of U is  $R^{\mathrm{U}}_{\mathrm{MT}}=2.79$~a.u..}
\label{fig:US-Fks}
\end{center}
\end{figure}

\begin{figure}[htbp]
\begin{center}
\includegraphics[width=0.4\columnwidth]{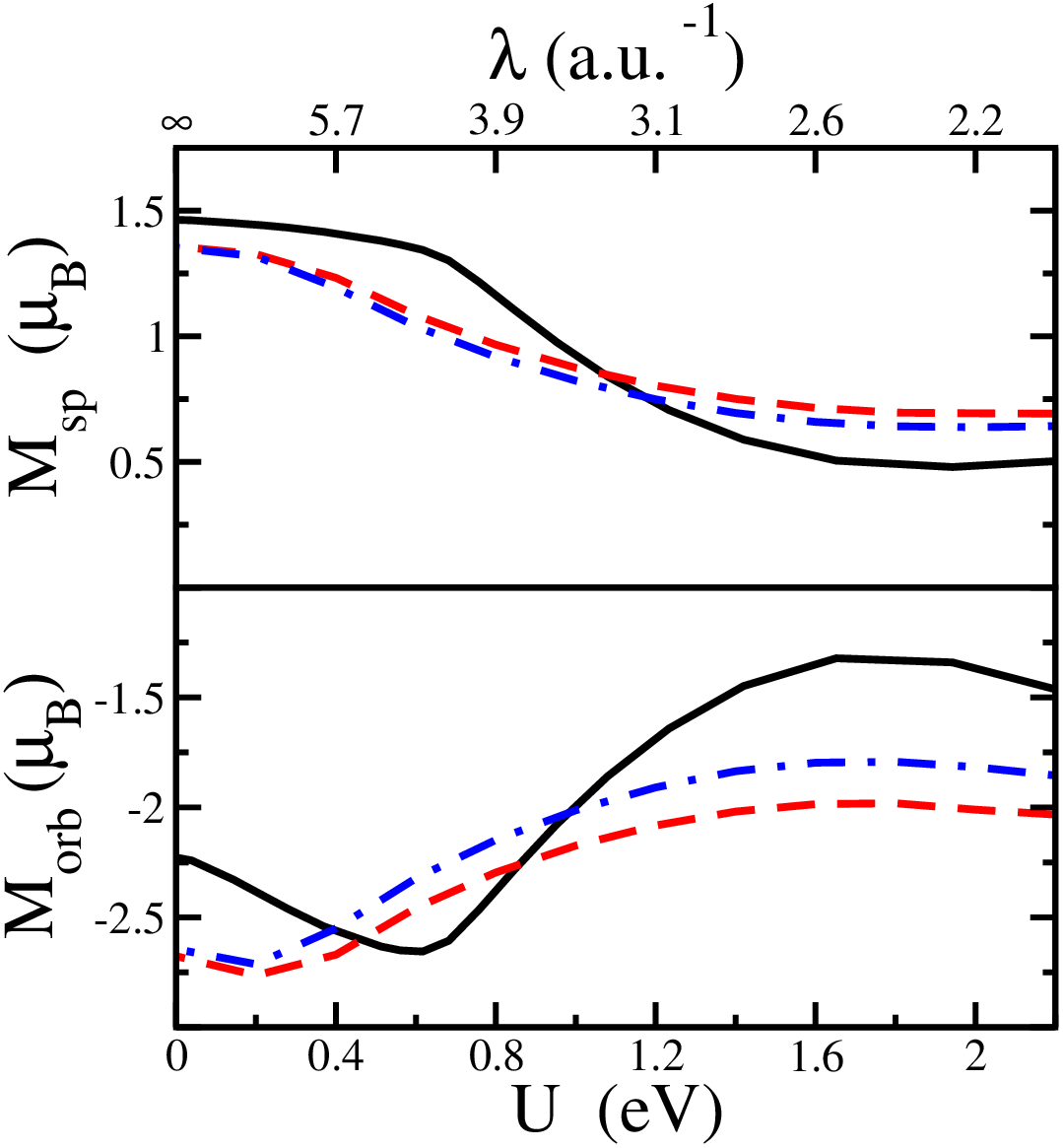}
\caption{Spin (upper panel) and orbital magnetic moment (lower panel) of US, $a=10.36$~a.u.,  calculated with LDA+SOC+$U$ approach and INT DC (see DC corrections in the method section). We compare the results obtained by calculating the $F^{(k)}$s with a screened Yukawa potential (full black line, screening length in the upper axis) with ones obtained by fixing the ratios $A_1$ and $A_2$ to ones of Ref.~\cite{Judd,LAZ} (dashed red line) and to ones of Ref.~\cite{Shishidou, Norman} (dashed-dotted blue line). In the fixed ratios calculations we fixed the parameter J=0.46~eV to one of Ref.~\onlinecite{Shishidou}. \\
The MT radii of U and S are set, respectively, to $R^{\mathrm{U}}_{\mathrm{MT}}=2.79$~a.u. and $R^{\mathrm{S}}_{\mathrm{MT}}=2.13$~a.u.. The parameter $R^{\mathrm{U}}_{\mathrm{MT}} |\vec{G}+\vec{k}|_{\mathrm{max}}$, governing  the number of plane waves in the APW+$lo$ method, is chosen to be 9.76. The BZ is sampled with 1728 $k$-points uniformly spaced. }
\label{fig:US-Mom-Jfix}
\end{center}
\end{figure}

Since in APW+$lo$ basis set  $R_{\ell}(r,\varepsilon)$ is energy dependent, we decided to use the energy $\varepsilon$ at the center of the band of the localized shell $\ell$. We set the atomic muffin-tin (MT) radius to a value large enough such that the integrals in Eq.~(\ref{eq:Fks}) are well converged.  
In the upper part of Fig.~\ref{fig:US-Fks} we plot the calculated Slater parameters for US. These values are in perfect agreement with ones calculated for the ion $U^{4+}$ by Norman~\cite{Norman} with the same screened potential. In the lower part of Fig.~\ref{fig:US-Fks} we compare the Slater integrals ratios $A_1$ and $A_2$ obtained for US with the fixed ratios commonly used in most LDA+$U$ studies~\cite{LAZ,Judd}. There is a good agreement only at small values of the screening length. For $\lambda \geq 0.5$~a.u.$^{-1}$ there start to be a significative difference that turns out to be relevant in the calculation of US magnetic moments. If the $F^{(k)}$s are calculated individually with Eq.~(\ref{eq:Fks}) and Eq.~(\ref{eq:Yukawa}), the spin moment ($M_\mathrm{sp}$) and orbital moment ($M_{\mathrm orb}$) show a more dramatic variation as a function of $U$ (or $\lambda$) than ones determined  by fixing $A_1$ and $A_2$ (see Fig.~\ref{fig:US-Mom-Jfix}). This fact indicates how relevant  the individual determination of every Slater integral might be in many systems.
Finally we verified that one can reproduces the LSDA moments in the limit of large screening length ($F^{(k)} \rightarrow 0$ for $\lambda \rightarrow \infty$), i.e. by slowly changing the parameter $\lambda$ one increases the localization of the $5f$ electrons from the LSDA limit in a continuos way.

A parameter-free method to screen the Slater parameter has been suggested by Brooks~\cite{Brooks_Fks}, where the screening parameter $\lambda$ is identified as the Thomas-Fermi screening, which depends on the local charge density. 
A more appropiate but time-consuming scheme to calculate the Slater parameters from the screened Coulomb potential is within the RPA approach, as it has been recently accomplished by Solovyev et al.~\cite{Solovyev-II}.

\subsection{Calculation of LDA+$U$ Potential}

The contribution to the orbital potential from the LDA+$U$ correction is defined as
\begin{equation}
V_{ij}=\frac{\delta E_{HF}}{\delta \rho_{ji}}=\sum_{ab} \left[\left<ja|g|ib\right> - \left<ja|g|bi\right>\right] \rho_{ab} \ .
\end{equation}
We note that the potential so defined is the complex conjugate of the one sometimes stated in literature.
The correct definition becomes crucial to evaluate off-diagonal spin terms in all calculations in which the coordinate system is not rotated to the local one, i.e. in all calculations in which the density matrix is not diagonal.

\subsection{Double-Counting Corrections}
A major obstacle in the LDA+$U$ approach is that the
electron-electron interaction has already been included in LDA potential, thus a simple addition of the orbital dependent HF potential would lead to DC terms. One may want to individuate those terms in the LDA potential that correspond to the interaction already considered in the HF Hamiltonian and subtract them. A direct connection between the two formalisms is not possible and in addition it would not be useful. In fact LDA approximation treats very accurately spatial variations of the exchange-correlation potential but it neglects  the orbital dependence of the Coulomb interaction. Thus the best recipe would be to identify the mean-field part of the HF potential and subtract it, leaving only an orbital dependent correction to the mean-field type LDA potential. Czy\.{z}yk and Sawatzky~\cite{Czyzyk} suggested a prescription that is exact in the case of uniform occupancies (around-mean-field, AMF) and that would be realistic for weakly correlated systems, however not exact because of the presence of the crystalline field. The AMF correction is implemented by redefining a new density matrix without the charge $n$ and the magnetization $\vec{m}$ contributions
\begin{eqnarray}
\label{eq:chg-mag}	
n &=& \mathrm{Tr} \rho \, \\
\vec{m} &=&  \mathrm{Tr} \, \vec{\sigma}\rho \ , 
\end{eqnarray}
in the following way
\begin{equation}
\tilde{\rho}_{ab}=
\rho_{ab}-(\delta_{ab}n+{\vec{\sigma}_{ab}\cdot\vec{m}})/{D}\ .
\end{equation}
The  AMF-double-counting corrected LDA+$U$ energy and potential terms become in our formalism
\begin{equation}
E_{HF-\mathrm{dc}}^\mathrm{AMF}=\frac{1}{2}\sum_{abcd} \tilde{\rho}_{ac} \left[\left<ab|g|cd\right> - \left<ab|g|dc\right>\right] \tilde{\rho}_{bd}\ ,
\end{equation}
\begin{equation}
V_{ij}^\mathrm{AFM}= V_{ij}  - \sum_{ab} \left[\left<ja|g|ib\right> - \left<ja|g|bi\right>\right] (\delta_{ab}n+{\vec{\sigma}_{ab}\cdot\vec{m}})/{D} = \sum_{ab} \left[\left<ja|g|ib\right> - \left<ja|g|bi\right>\right] \tilde{\rho}_{ab}\ .
\end{equation}

For strongly correlated systems it exists another prescription for the DC, the fully-localized-limit (FLL)~\cite{LAZ}, that would correspond to subtract  the average effect for a localized state, with integer occupation number. The most general expressions for energy and potential are 
\begin{equation}
E_\mathrm{dc}^\mathrm{FLL}=\left\{2Un(n-1)-2Jn(n/2-1)-J\vec{m}\cdot\vec{m}\right\}/4\ ,
\end{equation}
\begin{eqnarray}
V_{ij}^\mathrm{FLL}&=& V_{ij} - \left[\frac{U(2n-1)}{2}-\frac{J(n-1)}{2}\right]\delta_{ij}+\frac{J\vec{m}\cdot \vec{\sigma}_{ij}}{2}\ .
\end{eqnarray}  

Most of LDA+$U$ calculations use one of these approaches, while the real occupation numbers lie somewhere between the two limits.
 Petukhov et al.~\cite{Petukhov} proposed a linear interpolation between these two limits (INT DC),
\begin{equation}
E_{HF-\mathrm{dc}}^\mathrm{INT}=\alpha E_{U-\mathrm{dc}}^\mathrm{FLL}+(1-\alpha) E_{U-\mathrm{dc}}^\mathrm{AMF}\ ,
\label{eq:En-INT}
\end{equation}
\begin{equation}
V_{ij}^\mathrm{INT}=\alpha V_{ij}^\mathrm{FLL} + (1-\alpha) V_{ij}^\mathrm{AMF}\ ,
\label{eq:Pot-INT}
\end{equation}
 in which the parameter $\alpha$ is a material dependent constant determined in a self-consistent (SC) way following a constrained-DFT philosophy.  In our formalism the expression of $\alpha$ of Ref.~\cite{Petukhov} is generalized to take into account the off-diagonal spin terms of the density matrix, 
\begin{equation}
\alpha=\frac{D\,\mathrm{Tr}{\tilde{\rho} \cdot \tilde{\rho}}}{Dn- n^2-m^2} \ ,
\end{equation}
where $n$ and $m$ are defined in Eq.~(\ref{eq:chg-mag}).

In the present study we prefer to use the INT DC approach for two reasons.
Firstly, it reduces one further free parameter. The results do depend on the choice of DC, but if we stay consequently with the INT DC this degree of freedom is gone since we can in principle treat both more itinerant and more localized system.
Secondly, it turns out that the use of the INT DC is very important to reproduce the correct magnetic structure of  monopnictides Pu compounds. In Fig.~\ref{fig:PuS-PuP-Mom} we compare $M_{\mathrm{sp}}$ and $M_{\mathrm{orb}}$ calculated for the ferromagnetic PuS~\cite{HB} with ones calculated for the paramagnet PuP~\cite{HB}. For AMF type of DC, $M_{\mathrm{sp}}$ and $M_{\mathrm{orb}}$ of both compounds decrease dramatically until they disappear at $U\approx 2.0$~eV. Instead, by using FLL DC, $M_{\mathrm{sp}}$ and $M_{\mathrm{orb}}$ decrease significantly but they never disappear for any value of $U$. Finally, only by using the INT type of DC, we find a range of values for U ($U \gtrsim 4.0$~eV) in which PuS is magnetic and PuP is non-magnetic. 

 In conclusion we have implemented the LDA+$U$ method in the most general form, taking into account the off-diagonal spin terms of the density matrix and the correct definition of the potential for those terms.  By using the interpolated DC of Petukhov et al.~\cite{Petukhov} and by calculating SC the $F^{(k)}$s with a Yukawa potential, our LDA+$U$ approach has only one free parameter left, i.e.\ the screening length $\lambda$ or if preferable $U$.

\subsection{Multipole Representation of LDA+$U$ Energy}
\label{tensor}
The formalism up to now is standard and have been used several times before, although some 
studies have neglected the spin-mixing terms in Eq.~(\ref{eq:HF}). Within this formalism the density matrix plays a crucial role. In the following we will decompose this fairly large matrix into the most important and physical relevant terms. We will find that this decomposition largely simplifies the analysis, as well as it gives many new insights into the magnetism of the actinides, where the SOC has a crucial contribution.
The interaction term in Eq.~(\ref{eq:HF}) have been studied in detail, for example by Slater~\cite{Slater}, Racah~\cite{Racah-I,Racah-II,Racah-III,Racah-IV} Condon and Shortly~\cite{CS}. By expanding the interaction in spherical harmonics and by making use of the Wigner $3j$-symbols~\cite{CS} the interaction can be expressed as
\begin{align}
\label{int3j}
\left<ab|g|cd\right>&=\delta_{s_a s_c}\delta_{s_b s_d}(2\ell+1)^2\sum_{k=0}^{2\ell}\sum_{q=-k}^k(-1)^{m_a+m_b+q}\\
&\times\threej{\ell}{k}{\ell}{0}{0}{0}^2\threej{\ell}{k}{\ell}{-m_a}{-q}{m_c}F^{(k)}\threej{\ell}{k}{\ell}{-m_b}{q}{m_d}
\ .
\end{align}
The spin dependence is given by the two delta functions of the spin quantum numbers of states $a$, $b$, $c$ and $d$. The radial dependence is confined in the Slater integrals $F^{(k)}$ and the Wigner $3j$-symbols take care of the angular part of the integral. We now introduce a multipole momentum tensor $\mathbf{w}^k$ defined as the expectation values of a tensor operator $\mathbf{v}^k$,
\begin{align}
w^k_x&=\mathrm{Tr}\, v^k_x \rho \ , \\
v^k_x&\equiv\left<m_b| v^k_x|m_a\right>=(-1)^{\ell-m_b}\threej{\ell}{k}{\ell}{-m_b}{x}{m_a}n^{-1}_{\ell k} \ , \\
n_{\ell k}&=\frac{(2\ell)!}{\sqrt{(2\ell-k)!(2\ell+k+1)!}}\ , 
\end{align}
where the tensor component index $x$ runs from $-k$ to $k$.
The spin independent part of the HF energy (the Hartree term) can be rewritten as
\begin{align}
E_{H}&=\frac{(2\ell+1)^2}{2}\sum_{k=0}^{2\ell} n^2_{\ell k}\threej{\ell}{k}{\ell}{0}{0}{0}^2F^{(k)}\mathbf{w}^k\cdot\mathbf{w}^k\, .
\end{align}
In order to take care of the spin dependence we may introduce a double tensor
\begin{align}
\label{dt}
w^{kp}_{xy}&
=\mathrm{Tr}\, v^k_x t^p_y\rho \, \\
t^p_y&=(-1)^{s-s_b}\threej{s}{p}{s}{-s_b}{y}{s_a}n^{-1}_{sp}\ ,
\end{align}
where the index $y$ runs from $-p$ to $p$.
It is easy to verify that $\mathbf{w}^{k0}=\mathbf{w}^{k}$.
Then the exchange energy can also be written  as a function of the tensor components rather than the density matrix (see Appendix),
\begin{align}
\nonumber E_{\mathrm{X}}&=-\sum_{k=0}^{2\ell} F^{(k)}\sum_{k_1=0}^{2\ell}\frac{(2\ell+1)^2(2k_1+1)}{4}(-1)^{k_1}n^2_{\ell k_1}
\label{X-6j}\\
&\times\threej{\ell}{k}{\ell}{0}{0}{0}^2\sixj{\ell}{\ell}{k_1}{\ell}{\ell}{k}\sum_{p=0}^1\mathbf{w}^{k_1p}\cdot\mathbf{w}^{k_1p}\,,
\end{align}
where the $\{...\}$ symbol is the Wigner $6j$-symbol.
Notice that the Wigner $3j$-symbols are defined such that the contribution from odd $k$ vanish, so only Slater parameters of even $k$ are needed.
This type of expression was derived by Racah~\cite{Racah-II}. However, since it was derived for atomic configurations only, it has not been fully realised that it is as valid for non-integer occupations.

\subsection{The coupling of indices -- irreducible spherical tensor}

It is useful to introduce the irreducible spherical tensors $\mathbf{w}^{kpr}$ from the double tensors $\mathbf{w}^{kp}$ for two reasons. Firstly, the double tensors are not true spherical tensors and, secondly, in the presence of SOC the spin and orbital degrees of freedom are not longer decoupled.
The three-index tensors $\mathbf{w}^{kpr}$ is defined through a coupling of the indices of the double tensor
$\mathbf{w}^{kp}$,
\begin{align}
w^{kpr}_{t}&=\underline{n}_{kpr}^{-1}  \sum_{xay} (-)^{k-x+p-y} 
 \threej{k}{r}{p}{-{x}}{t}{-{y}}w^{kp}_{xy}\label{2-to-3} \ ,
\end{align}
where the index $r$ runs from $|k-p|$ to $|k+p|$ and where the normalization factor $\underline{n}_{abc}$ is given, as in Ref.~\cite{vdLaan1}, by
\begin{align}
\underline{n}_{abc}=i^g \left[\frac{(g-2a)!(g-2b)!(g-2c)!}{(g+1)!}\right]^{1/2}\frac{g!!}{(g-2a)!!(g-2b)!!(g-2c)!!} \ ,
\end{align}
with $g=a+b+c$.

These tensor moments have a very nice feature, they are proportional to the moment expansions of the charge ($k$ even and $p=0$), spin magnetization ($k$ even and $p=1$), current ($k$ odd and $p=0$) and spin current ($k$ odd and $p=1$) densities. For instance, $\mathbf{w}^{000}$ gives the total charge, $\mathbf{w}^{011}$ gives the spin moment and $\mathbf{w}^{101}$ is proportional to the orbital moment.

The exchange energy $E_{X}$ of the shell $\ell$ in terms of the irreducible spherical tensor moments is now
\begin{align}
\nonumber E_{X}&=-\sum_k F^{(k)}\sum_{k_1pr}\sum_{k_1 p r}\frac{(2\ell+1)^2(2k_1+1)(2r+1)}{4}(-1)^{k_1}|\underline{n}_{k_{1}pr}|^2n^2_{\ell k_1}
\\
&\times\threej{\ell}{k}{\ell}{0}{0}{0}^2\sixj{\ell}{\ell}{k_1}{\ell}{\ell}{k}\mathbf{w}^{k_1pr}\cdot\mathbf{w}^{k_1pr}\,.\label{eq:Ex-channel}
\end{align}

It is convenient to rewrite Eq.~(\ref{eq:Ex-channel}) in a simplified form,
\begin{align}
E_{\mathrm X}=\sum_{k_1pr} E_{\mathrm X}^{k_1pr}=\sum_{k_1pr} K_{k_1pr} \mathbf{w}^{k_1pr}\cdot\mathbf{w}^{k_1pr}\ ,
\label{eq:Ex_kpr}
\end{align}
where
\begin{align}
K_{k_1pr} = - \sum_{2k=0}^{4\ell} & F^{(k)}\frac{(2\ell+1)^2(2k_1+1)(2r+1)}{4}(-1)^{k_1}|\underline{n}_{k_{1}pr}|^2n^2_{\ell k_1} \\
& \times\threej{\ell}{k}{\ell}{0}{0}{0}^2\sixj{\ell}{\ell}{k_1}{\ell}{\ell}{k} \ .
\end{align}

In Eq.~(\ref{eq:Ex_kpr}) the exchange energy of the shell $\ell$ is expressed as a sum of independent terms involving different spherical tensors. We will refer to these terms as different exchange {\em channels}.

One of the most important types of polarization is the so-called spin polarization (SP), often referred to as Stoner exchange or Hund's first rule, which corresponds to a polarization of channel $011$. Since $\mathbf{w}^{011}\cdot\mathbf{w}^{011}=m_\mathrm{spin}^2$, we get that the SP energy $E_\mathrm{SP}$ is given by,
\begin{align}
E_\mathrm{SP}=E^{011}_X=K_{011}m_\mathrm{spin}^2=-\frac{1}{4}\left(\frac{U-J}{2\ell+1}+J\right)m_
\mathrm{spin}^2 \ .
\end{align}
Hence, the so-called Stoner parameter $I$, defined by $E_{\mathrm SP}=-\frac{1}{4}Im_\mathrm{spin}^2$, is given by
\begin{align}
I=\frac{U-J}{2\ell+1}+J \ , 
\end{align}
as it is already known.
In this expressions we have adopted the convention to use certain linear combinations of the Slater parameters, the ``Hubbard''-parameters, $U=F^{(0)}$, and $J$, that for $d$ and $f$ electrons is given by, respectively,
\begin{align}
J^{d}&=\frac{1}{14} (F^{(2)} + F^{(4)})  \ , \\
J^{f}&=\frac{2}{45}F^{(2)}+\frac{F^{(4)}}{33}+\frac{50}{1287}F^{(6)}  \ .
\end{align} 
In our multipole expansion in Eq.~(\ref{eq:Ex_kpr}) it is also included an exact formulation of the orbital polarization (OP) exchange energy $E_\mathrm{OP}$,
\begin{align}
E_\mathrm{OP}=E^{101}_X+E^{110}_X=K_{101}\mathbf{w}^{101}\cdot\mathbf{w}^{101}+ K_{110}\mathbf{w}^{110}\cdot\mathbf{w}^{110} \ ,
\end{align}
where $K_{101}=3K_{110}$ and $\mathbf{w}^{101}\cdot\mathbf{w}^{101}=\frac{m_\mathrm{orb}^2}{\ell^2}$.
This expression is a sum of two terms, one that breaks the time-reversal symmetry~\cite{parity}, 101 (OP-odd), and a second one that does not, 110 (OP-even). Consequently 101 is associated with the presence of an orbital moment, while 110 is compatible with a non-magnetic solution~\cite{Cricchio}. 
Finally, the prefactor $K_{101}$ has a simple expression in terms of Racah parameters~\cite{Judd}; this is for $d$ and for $f$ electrons, respectively,
\begin{align}
K^{d}_{101}&=-\frac{E^0+21E^2}{10} \ , \\
K^{f}_{101}&=-\frac{9E^0+297E^3}{112}  \ .
\end{align}

\section{Results}
In this section we shall apply our method to calculate the magnetic structure of some metallic U, Np and Pu compounds, for which the behavior of $5f$ electrons varies from itinerant to more localized.
\subsection{Application to US}
\begin{figure}[htbp]
\begin{center}
\includegraphics[width=0.45\columnwidth]{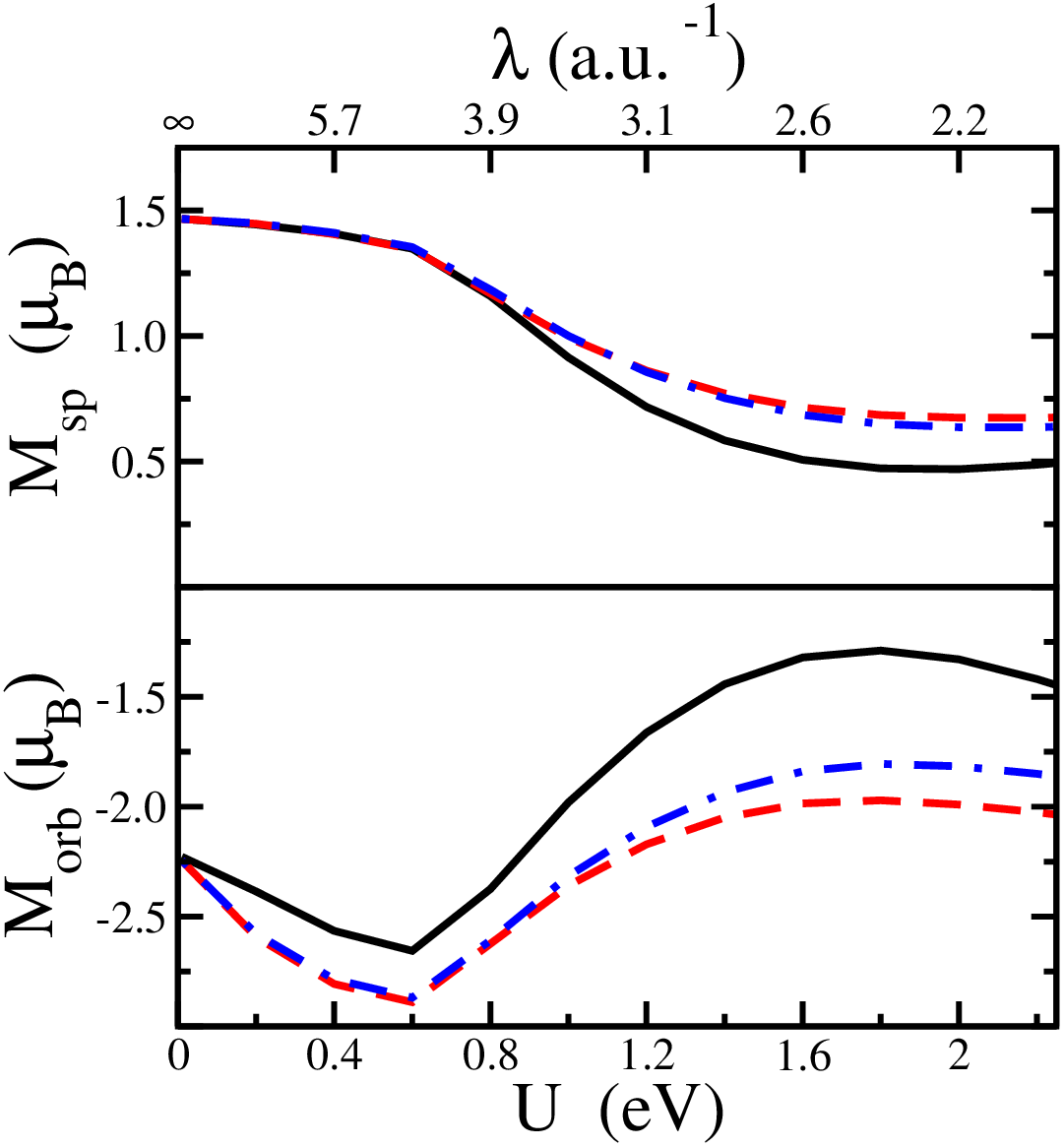}
\caption{Spin (upper panel) and orbital magnetic moment (lower panel) of US, $a=10.36$~a.u., calculated with LDA+SOC+$U$ approach. We compare the results obtained by calculating the $F^{(k)}$s with a screened Yukawa potential (full black line, screening lenght in the upper axis) with ones obtained by fixing the ratios $A_{1}=F^{(4)}/F^{(2)}$ and $A_{2}=F^{(6)}/F^{(2)}$ to ones of Ref.~\onlinecite{LAZ,Judd} (dashed red line) and to ones of Ref.~\onlinecite{Shishidou,Norman} (dashed-dotted blue line). In the fixed ratios calculations we varied $J$ to the one determined by Yukawa potential at every different screening length, $J(\lambda)$. Other calculation details are the same as the ones reported in Fig.~\ref{fig:US-Mom-Jfix}.}
\label{fig:US-Mom-Jvar}
\end{center}
\end{figure}

\begin{figure}[htbp]
\begin{center}
\includegraphics[width=0.5\columnwidth]{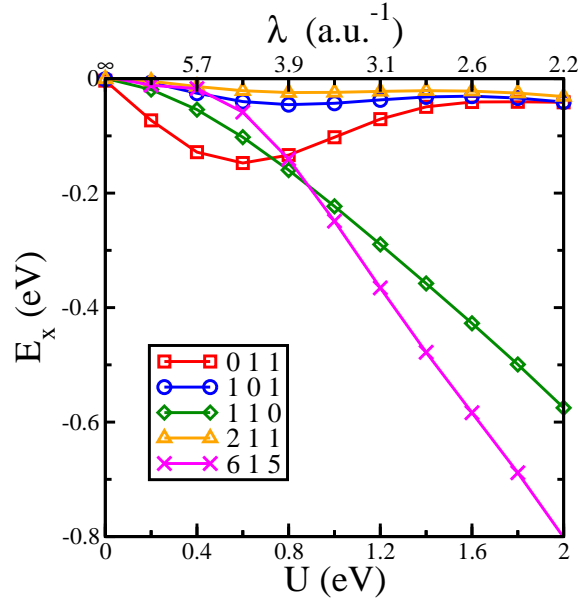}
\caption{Relevant exchange energy channels in Eq.~(\ref{eq:Ex_kpr}), $E_\mathrm{X}^{k_1pr}$,  of US  calculated with LDA+SOC+$U$ method and INT DC. The Slater parameters are calculated SC by using a screened Yukawa potential with screening length $\lambda$ reported in the upper axis. Other calculation details are reported in Fig.~\ref{fig:US-Mom-Jfix}.}
\label{fig:US-Ex}
\end{center}
\end{figure}

We selected US as a prototype system to compare the results obtained by calculating the $F^{(k)}$s through a screened Yukawa potential with other common procedures~\cite{LAZ, Shishidou}, for which $U$ and $J$ are provided as inputs and the ratios
$A_{1}=F^{(4)}/F^{(2)}$ and $A_{2}=F^{(6)}/F^{(2)}$ are fixed. US shows a ferromagnetic order with $T_C = 178$~K and a strong anisotropy along [111]  direction~\cite{US-111-I,US-111-II}. Many experiments have proven $5f$ electrons of US to be itinerant~\cite{US-It-I,US-It-II, US-It-III}. Neutron scattering experiments have measured the $5f$ total moment of US to be 1.7~$\mu_B$~\cite{US-Mom-5f}, while measurements of the magnetization in the bulk have shown that the total moment per formula unit is 1.55~$\mu_B$~\cite{US-Mom}. On the theoretical side there have been many investigations of magnetic properties of US, making this system a good benchmark case for our method. Recent LSDA calculation with SOC correction seem to underestimate the orbital moment~\cite{Brooks}, while orbital polarization~\cite{Brooks} and Hartree-Fock tight-binding calculations reproduce the correct size of $M_{\mathrm{orb}}$~\cite{Shishidou}. 
We have investigated the magnetic and electronic structure of this compound using our optimized LDA+SOC+$U$ method and its multipole decomposition. As anticipated in the method section, we compared the behavior of  magnetic moments  obtained by calculating $F^{(k)}$s through a Yukawa potential ($M^{\mathrm{Yukawa}}_{\mathrm{sp}}(\lambda)$ and  $M^{\mathrm{Yukawa}}_{\mathrm{orb}}(\lambda)$) with ones determined by fixing the ratios $A_1$ and $A_2$. In the fixed ratios calculations, firstly, we fixed the value of $J$ to 0.46~eV~\cite{Shishidou} for any values of $\lambda$ (Fig.~\ref{fig:US-Mom-Jfix}),  secondly, we varied $J$ as function of $\lambda$ to the value $J({\lambda})$ determined by the screened calculation (Fig.~\ref{fig:US-Mom-Jvar}).  If the $F^{(k)}$s are determined individually  $M^{\mathrm{Yukawa}}_{\mathrm{sp}}(\lambda)$ and  $M^{\mathrm{Yukawa}}_{\mathrm{orb}}(\lambda)$ change dramatically as function of $\lambda$ (or $U$), while the moments are much more constant by fixing $A_1$, $A_2$ and $J$. If then $J$ is varied to $J({\lambda})$, and only  $A_1$, $A_2$ are kept constant, the moments vary in a similar way to $M^{\mathrm{Yukawa}}_{\mathrm{sp}}(\lambda)$ and  $M^{\mathrm{Yukawa}}_{\mathrm{orb}}(\lambda)$. This because in the latter case the values of $F^{(k)}$s become fairly similar to ones calculated with the screened potential, as we show in Tab.~\ref{tab:US-Fks}.\\
Let us now analyze the decomposition in multipoles of the HF-$E_\mathrm{X}$ energy in Eq.~(\ref{eq:Ex_kpr}), that we report in Fig.~\ref{fig:US-Ex}. We found three main contributions: the SP term 011, the OP-even term 110 and finally the high multipole of $\vec{m}(\vec{r})$ 615.
For small $U$ the SP 011 is the dominant contribution until it starts to rapidly decrease. At $U \approx 0.6$~eV the OP-even 110 crosses 011 and it becomes the largest term. Such a term has been detected in all our LDA+SOC+$U$ calculations on actinides and it has been found responsible for the vanishing of magnetic moments in $\delta$-Pu both in a LDA+$U$ investigation~\cite{Cricchio} as well as in DMFT study~\cite{Granas}. Finally at $U \approx 0.6$~eV the 615 contribution becomes the largest. 

Few other terms are present but they are much smaller than the others; 101 , OP-odd term, that refers to the $E_\mathrm{X}$ associated to the presence of an orbital moment, 211 and 413, related to the multipoles of the $\vec{m}(\vec{r})$, 505, corresponding to an high multipole of the current density. 

\subsection{Application to Np compounds}
\begin{figure}[htbp]
\begin{center}
\includegraphics[width=0.45\columnwidth]{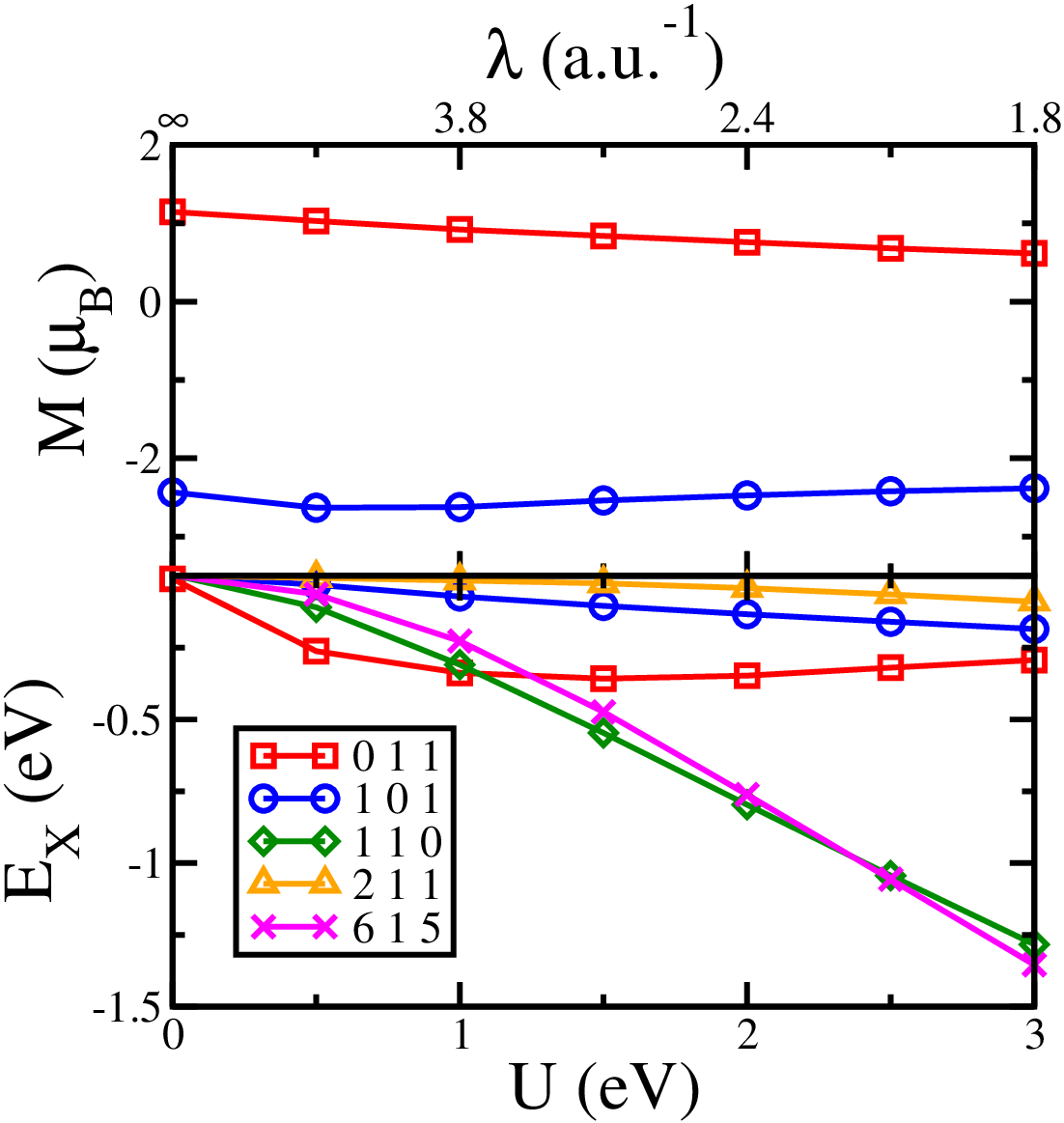}
\caption{Upper panel: spin (red line with squares) and orbital (blue line with circles) moment of ferromagnetic NpN, $a=9.25$~a.u.. 
Lower panel: relevant exchange energy channels in Eq.~(\ref{eq:Ex_kpr}), $E_\mathrm{X}^{k_1pr}$, of NpN calculated with LDA+SOC+$U$ approach and INT DC. The Slater parameters are calculated SC by using a screened Yukawa potential with screening length $\lambda$ reported in the upper axis. The MT radii of Np and N are set, respectively, to 
$R^{\mathrm{Np}}_{\mathrm{MT}}=2.60$~a.u. and $R^{\mathrm{N}}_{\mathrm{MT}}=1.80$~a.u.; the parameter $R^{\mathrm{Np}}_{\mathrm{MT}} |\vec{G}+\vec{k}|_{\mathrm{max}}$ is set to 9.4.  
The integration in the BZ is performed with 1728 $k$-points uniformly spaced.}
\label{fig:NpN-Mom-Ex}
\end{center}
\end{figure}

\begin{figure}[htbp]
\begin{center}
\includegraphics[width=0.45\columnwidth]{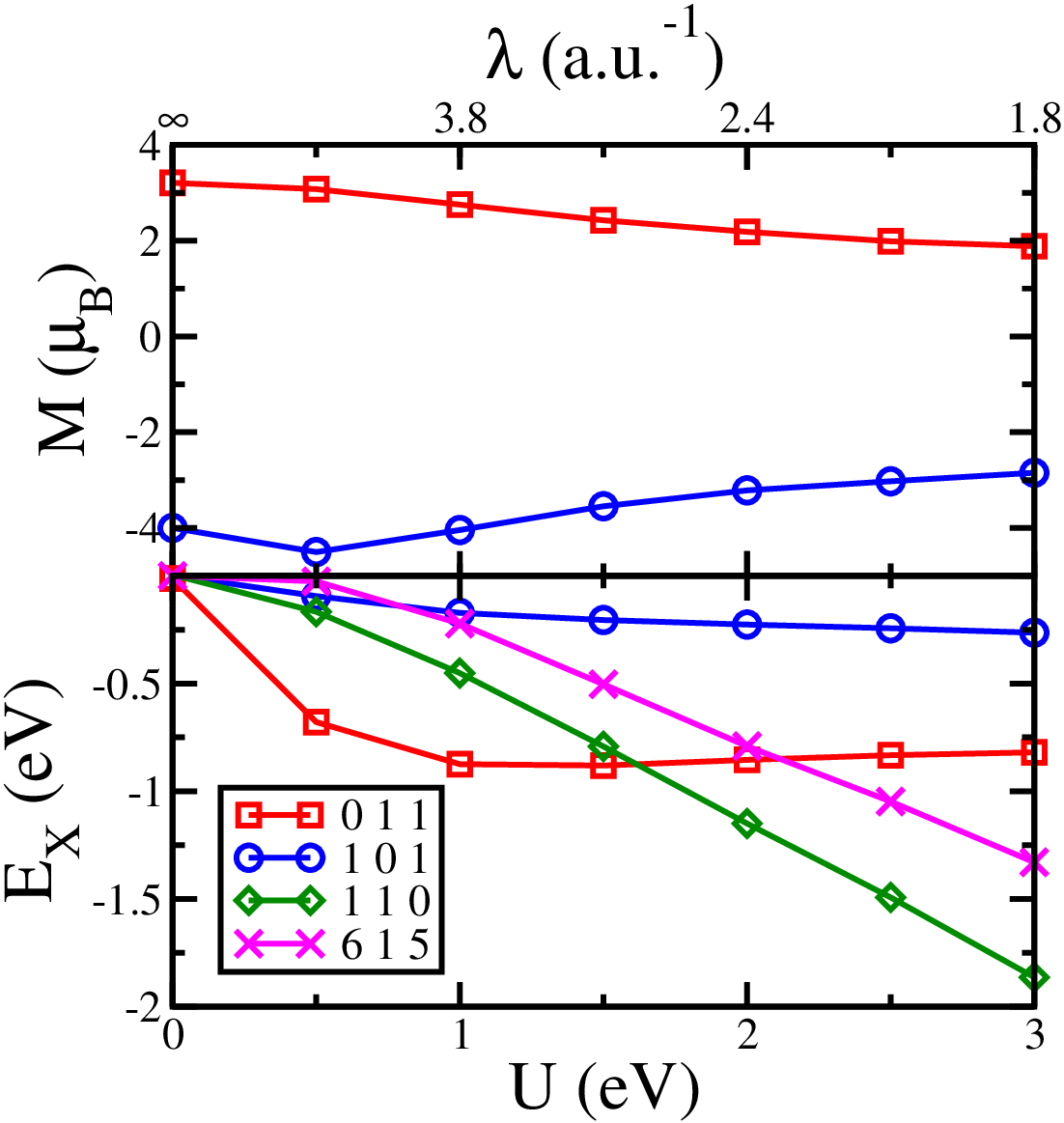}
\caption{Upper panel: spin (red line with squares) and orbital (blue line with circles) moment of NpSb in AFM-3k structure, $a=11.82$~a.u.. Lower panel: relevant  exchange energy channels in Eq.~(\ref{eq:Ex_kpr}), $E_\mathrm{X}^{k_1pr}$, of NpSb calculated with LDA+SOC+$U$ and INT DC. The Slater parameters are calculated SC by using a screened Yukawa potential with screening length $\lambda$ reported in the upper axis.
The MT radii of Np and Sb are set, respectively, to 
$R^{\mathrm{Np}}_{\mathrm{MT}}=2.95$~a.u. and $R^{\mathrm{Sb}}_{\mathrm{MT}}=2.65$~a.u.; the parameter $R^{\mathrm{Np}}_{\mathrm{MT}} |\vec{G}+\vec{k}|_{\mathrm{max}}$ is set to be 9.2.  
The integration in the BZ is performed with 1728 $k$-points uniformly spaced.}
\label{fig:NpSb-Mom-Ex}
\end{center}
\end{figure}

The $5f$ electrons in Np compounds are expected to have intermediate features between the itinerant behavior in U compounds and the more localized behavior in Pu compounds.
We calculated the electronic and magnetic structure of two Np compounds, NpN and NpSb. NpN is reported to be a ferromagnet with easy axis along [111] direction, $T_C=87~K$ and a total magnetic moment of $1.4~\mu_B$~\cite{HB}, NpSb is reported to be an anti-ferromagnet in AFM-3k structure and easy axis along [111] direction with a total moment of  $2.5~\mu_B$~\cite{HB}.
In Fig.~\ref{fig:NpN-Mom-Ex} and Fig.~\ref{fig:NpSb-Mom-Ex} we plot $M_\mathrm{sp}$ and $M_\mathrm{orb}$ as function of the parameter $U$ and the corresponding screening length $\lambda$. Both compounds are magnetic for all values of $\lambda$. In the same figures we also report the different multipole components of $E_\mathrm{X}$ as defined in Eq.~(\ref{eq:Ex_kpr}). For $0 \leq U \lesssim 1.5$~eV the SP 011 term dominates in both compounds. For $U \gtrsim1.5$~eV, 110 and 615 become the largest terms. In both materials there is also a significant  OP-odd 101, that is related to the presence of an orbital moment. Finally, in NpN we observe a significant 211 contribution, that corresponds to an high multipole of $\vec{m}(\vec{r})$. We note that the relevant channels in these Np compounds are similar to ones present in US, however the 110 term is larger and the 615 term is slightly smaller.

\subsection{Application to Pu compounds}
\begin{figure}[htbp]
\begin{center}
\includegraphics[width=0.75\columnwidth]{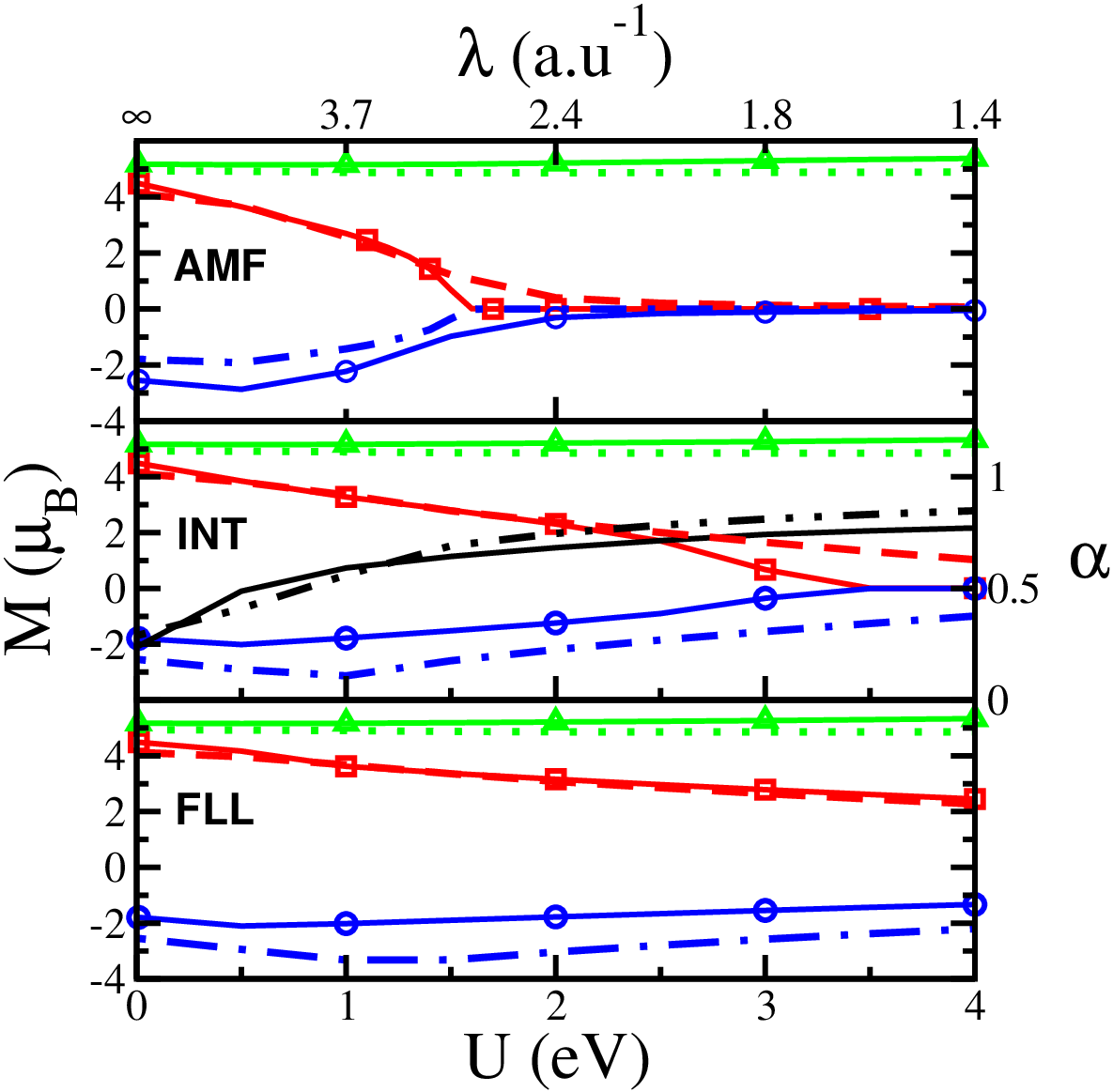}
\caption{ Spin $M_{\mathrm{sp}}$ (red lines) and orbital $M_{\mathrm{orb}}$ (blue lines) magnetic moments of paramagnetic PuS (full line with squares for $M_{\mathrm{sp}}$, full line with circles for $M_{\mathrm{orb}}$), $a=10.46$~a.u., and ferromagnetic PuP (dashed line for $M_{\mathrm{sp}}$, dashed-dotted line for $M_{\mathrm{orb}}$), $a=10.49$~a.u., as function of $U$ for different types of DC. 
The Slater parameters are calculated SC by using a Yukawa
potential with screening length $\lambda$ reported in the upper axis. For AMF DC the
moments vanish in both compounds at $U \approx 2.0$ eV, for FLL DC both compounds stay magnetic for all values of $U$.
Only for the INT DC there is a range of U ($U \gtrsim 4.0$~eV) for which we obtain the
experimental magnetic structure of both compounds~\cite{HB}; PuS becomes non-magnetic while PuP
stays magnetic. The green lines refer to the $5f$ charge of PuS (full line with triangles) and PuP (dotted line). The black lines refer to INT DC factor $\alpha$ of PuS (full line) and PuP (dashed-two-dots line).
The MT radii of Pu and S(P) are set, respectively, to 
$R^{\mathrm{Pu}}_{\mathrm{MT}}=2.82$~a.u. and $R^{\mathrm{S(P)}}_{\mathrm{MT}}=2.15$~a.u.; the basis set cut-off $R^{\mathrm{Pu}}_{\mathrm{MT}} |\vec{G}+\vec{k}]_{\mathrm{max}}$ is set to be 9.1.  
The integration in the BZ is performed with 1728  with $k$-points uniformly spaced.}
\label{fig:PuS-PuP-Mom}
\end{center}
\end{figure}

\begin{figure}[htbp]
\begin{center}
\includegraphics[width=0.45\columnwidth]{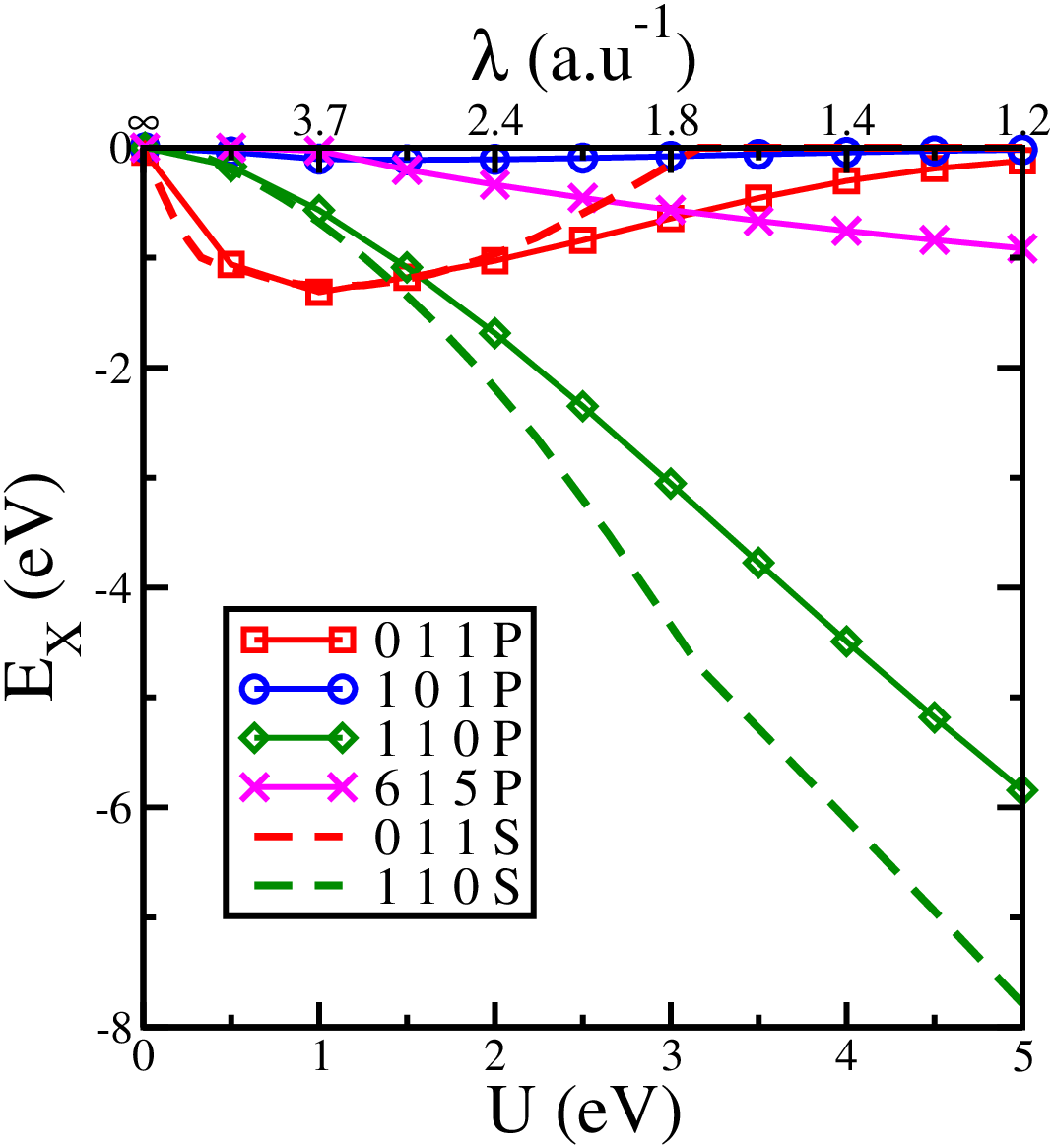}
\caption{Relevant exchange energy channels in Eq.~(\ref{eq:Ex_kpr}), $E_\mathrm{X}^{k_1pr}$, for paramagnetic PuS, $a=10.46$~a.u., and ferromagnetic PuP, $a=10.49$~a.u., as function of $U$ for INT DC. The Slater parameters are calculated SC by using a Yukawa
potential with screening lenght $\lambda$ reported in the upper axis. Other calculation details are reported in Fig.~\ref{fig:PuS-PuP-Mom}.}
\label{fig:PuS-PuP-Ex}
\end{center}
\end{figure}

\begin{figure}[htbp]
\begin{center}
\includegraphics[width=0.45\columnwidth]{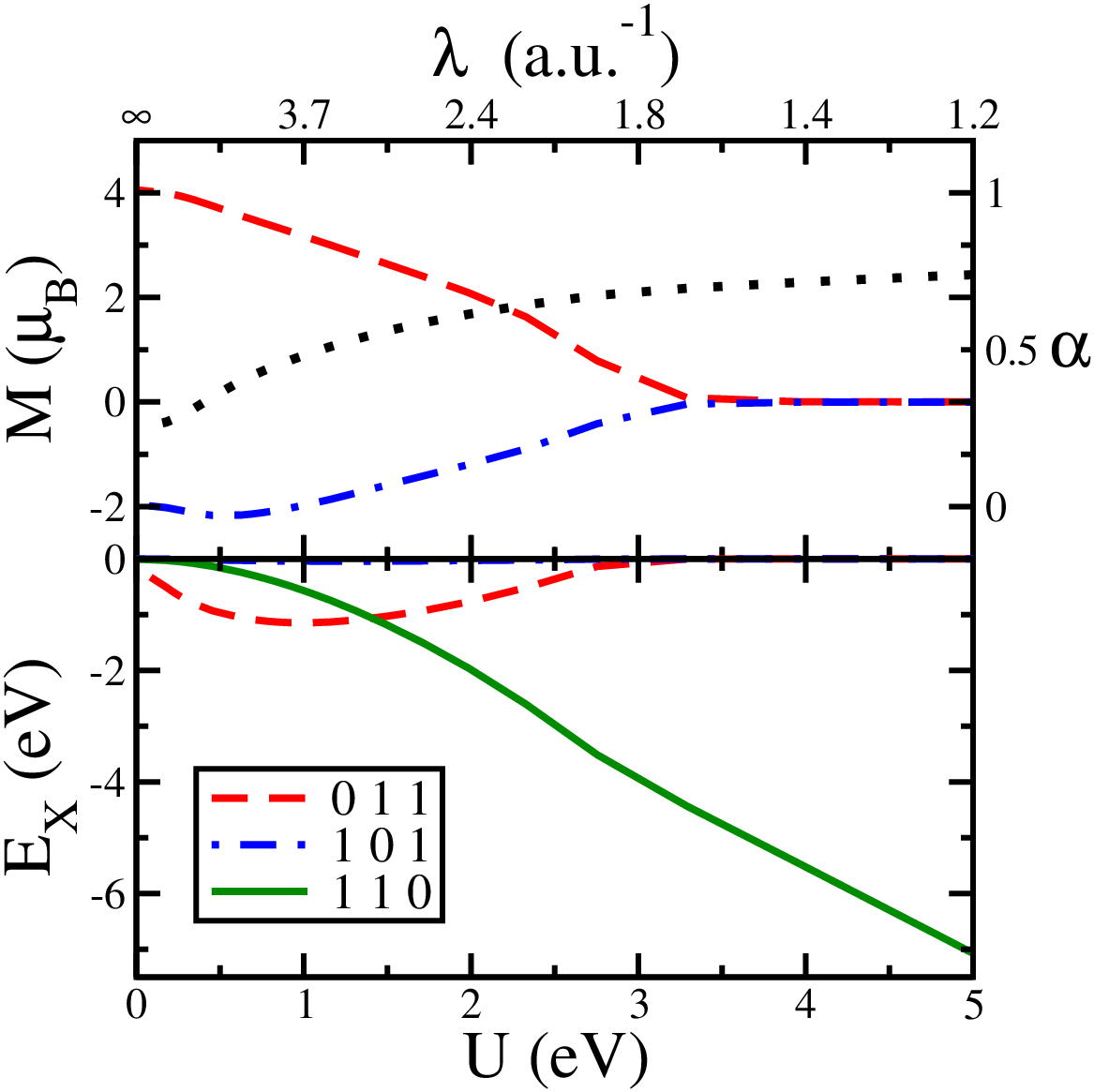}
\caption{Upper panel: spin (dashed red line) and orbital (dashed-dotted blue line) magnetic moments of high-$T_c$ superconductor PuCoGa$_{5}$, $a=7.84$~a.u and $c=12.57$~a.u., calculated with LDA+SOC+$U$ and INT DC, as function of parameter $U$. The interpolation factor $\alpha$ is plotted with a dotted black line. The Slater parameters are calculated SC by using a Yukawa
potential with screening lenght $\lambda$ reported in the upper axis. We used an AFM-1k structure with $\mathbf{q}=(0,0,1/2)$. Lower panel: relevant exchange energy channels in Eq.~(\ref{eq:Ex_kpr}), $E_\mathrm{X}^{k_1pr}$, of PuCoGa$_{5}$.\ \
The muffin-tin (MT) radii of Pu, Co and Ga are set, respectively, to 
$R^{\mathrm{Pu}}_{\mathrm{MT}}=2.7$~a.u. and $R^{\mathrm{Co,Ga}}_{\mathrm{MT}}=$2.2~a.u.. The parameter $R^{\mathrm{Pu}}_{\mathrm{MT}} |\vec{G}+\vec{k}|_{\mathrm{max}}$ is set to 9.5.  
The integration in the BZ is performed with 576 $k$-points uniformly spaced.}
\label{fig:PuCoGa5-Mom-Ex}
\end{center}
\end{figure}

 We have investigated the magnetic structure of some Pu compounds belonging to the actinides monopnictides, whose $5f$-electrons are expected to have more localized features compared to Np and U compounds. These materials have the fcc NaCl crystal structure and the determination of their magnetic structure has been object of many experimental studies~\cite{HB}. PuS, PuSe and PuTe, are paramagnets like $\delta$-Pu; PuSb, PuP, PuAs are ferromagnets; PuBi is an antiferromagnet.  We have found that the use of the INT DC is essential to correctly reproduce the magnetic structure of those compounds. In addition we reproduced the non-magnetic ground state of the high-$T_c$ superconductor PuCoGa$_{5}$ and we analyzed it through the multipole decomposition of $E_\mathrm{X}$ in Eq.~(\ref{eq:Ex_kpr}).  
We already discussed in the method section the ability of INT DC to correctly describe the magnetic structure of two prototype Pu compounds, the ferromagnet PuS and the paramagnet PuP (see Fig.~\ref{fig:PuS-PuP-Mom}). The results of a similar calculation for the  ferromagnets  PuAs, PuSb, for the antiferromagnet PuBi and for the paramagnets PuSe and PuTe are summarized in Tab.~\ref{Tab:Pu-Comp-Mom}. By using the  INT type of DC,  $M_{\mathrm{sp}}$ and $M_{\mathrm{orb}}$ decrease significantly faster for non-magnetic compounds until they disappear, while for magnetic compounds the moments decrease slower and do not disappear.\\
Let is now discuss the decomposition in multipoles of the HF-$E_\mathrm{X}$. Again we will refer to the two prototype materials PuS and PuP calculated with INT DC. In both compounds the dominant term is  SP 011 for  $0 \leq U \lesssim 1.5$~eV. With increasing $U$ the OP-even term 110 starts to increases and it soon becomes the dominant contribution, for $U \gtrsim  1.5$~eV. In PuS the OP-even 110 term completely takes over the SP 011 contribution, as we already found in the case of $\delta$-Pu~\cite{Cricchio}. Also in PuP 011 decreases, however it does not become zero and few other significant channels are opened. The most relevant one (after 110) is the 615 contribution that is related to an high multipole of $\vec{m}(\vec{r})$. However, the 615 term is lower than one present in magnetic U and Np compounds, while 110 channel is clearly larger.
This mechanism is clearly shown in Fig.~\ref{fig:PuS-PuP-Mom} and Fig.~\ref{fig:PuS-PuP-Ex}, where we compare the magnetic moment and HF-$E_\mathrm{X}$ contributions of PuS, non-magnetic, and PuP, ferromagnetic, as function of the parameter $U$ and corresponding screening length $\lambda$.\\
Finally we applied our one-parameter LDA+$U$ method to the superconducor PuCoGa$_{5}$.
Previous LDA+$U$ calculations of this compound have stabilized a non-magnetic solution with AMF type of DC, $U$=3~eV, $J$=0.6~eV and fixed $A_1$ and $A_2$~\cite{Shick_PuCoGa5}. We also stabilized  a non-magnetic solution for $U$=3.2 eV, corresponding to $\lambda=1.79$~a.u.$^{-1}$, (see Fig.~\ref{fig:PuCoGa5-Mom-Ex} and Table~\ref{Tab:Pu-Comp-Mom} ). In the $E_\mathrm{X}$ multipole decomposition the OP-even 110 term, that corresponds to an enhancement of SOC interaction, is again the dominant one taking over the whole SP 011 contribution once the system becomes non-magnetic. This mechanism is completely analogous to the one present in the others paramagnetic Pu monopnictides we calculated in this work and in $\delta$-Pu~\cite{Cricchio}. 

\section{Conclusions}
The purposes of this paper are two-fold. Firstly, we advocates an approach to LDA+$U$ calculations that reduces the number of free parameters in a well-defined way. This involves the use of screened Yukawa potential to describe the HF interaction within an atomic shell with a single parameter.
In addition we argue to systematically use the INT approach to DC, which takes away the ever existing choice between AMF or FLL, especially since the results depend on the choice of DC.
This approach ought to facilitate fast and systematic LDA+$U$ calculations. The results of  which would be comparable between different computations, without having to dwell on the used values of e.g.~$J$.

Secondly, we present a method, the decomposition in tensor moments of the density matrix, that facilitates the analyzes of the results from an LDA+$U$ calculation. This is an exact approach which gives both the different polarization channels as well as the corresponding Hartree and exchange energies. 

Finally,  we apply these combined approaches to a few metallic actinide systems whose $5f$ electrons show different degrees of localization; from the itinerant features in US to the more localized behavior in Pu compounds and the intermediate one of Np based materials. The results show that our one-parameter approach catches the ground state properties of these system, like spin and orbital moments. Here it is worth mentioning that an even better agreement with experiments can be obtained by tweaking the individual Slater parameters differently for the different systems. But that would rather reduce than increase  the physical understanding of these systems.

Most importantly, our results signal that these systems are not as well understood as believed.
For instance,  Hund's rules are far away from fulfilled, since the SP does not play the dominant role as always assumed.
On one hand the LDA calculations show large SP which leads to an overestimation of the spin moments.
On the other hand, when including the HF interaction of the LDA+$U$ functional, another polarization channel dominates, the one of spin-orbital currents, the OP-even 110 channel. The tensor moment associated with such channel, $ \mathbf{w}^{110}$, is even close to the saturation value of $-\frac{4}{3} \mathbf{w}^{000}$ for all systems considered. 
Large values of this tensor moment are in good accordance with the rising number of experiments
that point to large $\mathbf{w}^{110}$ for actinide systems~\cite{vdLaan2,Moore1,Moore2}.
This fact corresponds to a large enhancement of SOC in all these systems by the HF exchange interaction.
The SP 011-channel, however,  decreases drastically, and in some Pu compounds even disappears.
Instead, to some surprise, a third polarization plays a large role, the 615-channel; this corresponds to an high multipole of the magnetization density.

\section*{Acknowledgments}
The support from the Swedish Research Council (VR) is thankfully acknowledged. The computer calculations have been performed at the Swedish high performance centers HPC2N, NSC and UPPMAX under grants provided by the Swedish National Infrastructure for Computing (SNIC). 

\section{Appendix}
\label{Appendix}
The transformation Eq.~(\ref{dt}) can be reversed through orthogonality relations of the Wigner $3j$-symbols. Hence, the density matrix $\rho$ can be expanded in the double tensors $\mathbf{w}^{k_{i}p_{i}}$,
\begin{align}
\rho_{ac}&=\sum_{k_{i}x_{i}}(2k_{i}+1)n_{lk_{i}}(-1)^{m_{c}-\ell}\threej{\ell}{k_{i}}{\ell}{-{m}_{c}}{x_{i}}{m_{a}}\sum_{p_{i}y_{i}}(2p_{i}+1)n_{sp_{i}}(-1)^{s_{c}-s}\threej{s}{p_{i}}{s}{-{s}_{c}}{y_{i}}{s_{a}}w^{k_{i}p_{i}}_{x_{i}y_{i}}\label{tensor-to-dm}\, .
\end{align}
When this form, together with Eq.~(\ref{int3j}), are inserted in the exchange part of Eq.~(\ref{eq:HF}),
we get an expression where the complications essentially arise from the orbital summations in the factor
\begin{align}\label{Q}
{\cal Q}=\sum_{m_{a}m_{b}m_{c}m_{d}q}(-1)^q
\threej{\ell}{k_{1}}{\ell}{-{m}_{c}}{x_{1}}{m_{a}} 
\threej{\ell}{k}{\ell}{-{m}_{a}}{-{q}}{m_{d}} 
\threej{\ell}{k_{2}}{\ell}{-{m}_{d}}{x_{2}}{m_{b}}
\threej{\ell}{k}{\ell}{-{m}_{b}}{q}{m_{c}}\,,
\end{align}
where the indeces on $k$ and $x$ stem from the two different density matrix expansions.
The spin dependence of the exchange energy is simpler due to the Kronecker delta-symbols in Eq.~(\ref{int3j}), i.e. the relevant factor becomes
\begin{align}
\mathcal{S}=&\sum_{s_{a}s_{b}s_{c}s_{d}}(-1)^{-s_{c}-s_{d}}
\threej{s}{p_{1}}{s}{-{s}_{c}}{y_{1}}{s_{a}} 
\delta_{s_{a} s_{d}} 
\threej{s}{p_{2}}{s}{-{s}_{d}}{y_{2}}{s_{b}}
\delta_{s_{b} s_{c}}\\
=&
\sum_{s_{a}s_{b}}(-1)^{1+s_{a}-s_{b}}
\threej{s}{p_{1}}{s}{-{s}_{a}}{-y_{1}}{s_{b}} 
\threej{s}{p_{2}}{s}{-{s}_{a}}{y_{2}}{s_{b}}= (-1)^{1-y_{1}}(2p_{1}+1)^{-1}\delta_{p_{1} p_{2}}\delta_{y_{1} -y_{2}}
\label{S}\,,
\end{align}
by use of orthogonality relations.
To simplify $\mathcal Q$ we start with an identity for the Wigner $6j$-symbols, see e.g.\ Ref.~\onlinecite{Judd}, and we reshuffle a little,
\begin{align}
\sixj{\ell}{\ell}{k_{1}}{\ell}{\ell}{k}&
\threej{\ell}{\ell}{k_{1}}{m_{b}}{-{m}_{c}}{-{x}_{1}}
=\sum_{m_{a}m_{d}q}(-1)^{\ell+\ell+k+m_{a}+m_{d}+q}\\
\times&
\threej{\ell}{k}{\ell}{-{m}_{d}}{-{q}}{m_{b}}
\threej{\ell}{k}{\ell}{-{m}_{c}}{q}{m_{a}}
\threej{\ell}{k_{1}}{\ell}{m_{d}}{-{x}_{1}}{-{m}_{a}}\, .
\end{align}
Then we multiply by $\threej{\ell}{\ell}{k_{2}}{m_{b}}{-{m}_{c}}{x_{2}}$
and sum over $m_{b}$ and $m_{c}$, by obtaining
\begin{align}
&\sixj{\ell}{\ell}{k_{1}}{\ell}{\ell}{k}
\sum_{m_{b}m_{c}}
\threej{\ell}{\ell}{k_{1}}{m_{b}}{-{m}_{c}}{-{x}_{1}}
\threej{\ell}{\ell}{k_{2}}{m_{b}}{-{m}_{c}}{x_{2}}\\
&=\sum_{m_{a}m_{b}m_{c}m_{d}q}(-1)^{k_{1}+k+x_{1}+q}
\threej{\ell}{k_{1}}{\ell}{-m_{d}}{{x}_{1}}{{m}_{a}}
\threej{\ell}{k}{\ell}{-{m}_{a}}{-q}{m_{c}}
\threej{\ell}{k_{2}}{\ell}{-m_{c}}{x_{2}}{{m}_{b}}
\threej{\ell}{k}{\ell}{-{m}_{b}}{{q}}{m_{d}}
\, .
\end{align}
On the LHS we use an orthogonality relation for the Wigner $3j$-symbols and on the RHS we identify $\mathcal{Q}$ from Eq.~(\ref{Q})
\begin{align}
\label{Q6j}
\sixj{\ell}{\ell}{k_{1}}{\ell}{\ell}{k}&(2k_{1}+1)^{-1}\delta_{k_{1} k_{2}}
\delta_{x_{1} -x_{2}} = (-1)^{k_{1}+x_{1}+k}\, \mathcal{Q}\, .
\end{align}
The phase factors of Eqs.\ (\ref{S}) and (\ref{Q6j}) are then used to form the scalar product of the double tensors in the final form of exchange energy, Eq.~(\ref{X-6j}) as
\begin{align}
\mathbf{w}^{k_{1}p_{1}}\cdot\mathbf{w}^{k_{1}p_{1}}=\sum_{x_{1}y_{1}} (-1)^{x_{1}+y_{1}} {w}^{k_{1}p_{1}}_{x_{1}y_{1}}{w}^{k_{1}p_{1}}_{-x_{1}-y_{1}}\, .
\end{align}



\begin{table}
 \centering 
 \caption{Comparison of Slater integrals of US and $U^{4+}$ calculated with different methods. The fixed ratios $A_1$ and $A_2$ are from Ref.~\onlinecite{LAZ} and~\onlinecite{Judd}; in those calculations the parameters $U$ and $J$ are provided as input equal to ones calculated with Yukawa potential for $\lambda=1.6$~a.u.$^{-1}$. The Slater integrals of US are evaluated using the $5f$ radial functions with an energy 
 corresponding to the center of the band. The MT of U is 2.79~a.u..}
\begin{tabular}{|c|c|c|c|c|c|c|}
\hline                              
 &  Method                                                                & $F^{(0)}$        & $F^{(2)}$         & $F^{(4)}$      & $F^{(6)}$    & J \\
 \hline 
  Present Work US & $\lambda$=1.6 a.u.$^{-1}$                    &  3.114    &   6.128   &  5.110  &  4.060  &  0.585 \\
  Present Work US & Fixed $A_1$ and $A_2$ &   3.114   &   6.973  &   4.659  &  3.446  &  0.585  \\
  Ref.~\cite{Norman}  U$^{4+}$  & $\lambda$=1.6 a.u.$^{-1}$   &  3.309   &  6.377   &  5.281  & 4.185   &  0.606 \\
  Ref.~\cite{Brewer,Lang}   U$^{4+}$  & Experiment                 &  2.3, 2.6  & 6.440  & 5.296   & 3.441   &  0.580 \\
 \hline 
\end{tabular}
\label{tab:US-Fks}
\end{table}
\begin{table}
 \centering 
 \caption{Table of screening length $\lambda$ used to determine Slater integrals through a screened Yukawa potential in LDA+SOC+$U$ calculations of Pu compounds with INT type of DC. 
 We report the smallest value of $U$ and corresponding screening length $\lambda$ that are necessary to obtain vanishing magnetic moments (less than 0.1~$\mu_B$ for both $M_{\mathrm{sp}}$ and $M_{\mathrm{orb}}$) in non-magnetic Pu compounds. For the first three non-magnetic compounds PuX, with X=S,Se,Te, we considered the corresponding ferromagnetic compounds PuY, with Y=P,As,Sb,  where X and Y are chosen to be neighbours elements in the periodic table. We then report the magnetic moments of PuY using the same value of $U$ that makes the corresponding compound PuX non-magnetic. For the antiferromagnet PuBi we used the same value of $U$ of PuSb. We also write down the INT DC interpolation factor $\alpha$ that is determined SC. Values of experimental magnetic moments are from Ref.~\onlinecite{HB}.}
  \begin{tabular}{cccccccc}
\hline  \hline 
    Paramagnetic \\
\hline \\¤
                             & U~[eV]   &  $\lambda$~[a.u$^{-1}$]  & $\alpha$   \\
   PuS                   &   3.2                           &  1.74     &         0.75                \\
   PuSe                 &    3.6                          &  1.57     &        0.79                  \\
   PuTe                 &    4.1                           &  1.38      &        0.83                   \\
   $\delta$-Pu      &    3.7                           &  1.51      &       0.78               \\
   PuCoGa$_{5}$         &    3.2                           &  1.79       &      0.69                \\

\hline  \hline 
    Ferromagnetic \\
\hline \\
   &  U~[eV]   & $\lambda$~[a.u$^{-1}$]  &  $M_{\mathrm{spin}}~[\mu_B] $ & $M_{\mathrm{orb}}~[\mu_B] $ & $M_{\mathrm{tot}}~[\mu_B] $ &  $M^{\mathrm{exp}}_{\mathrm{tot}}~[\mu_B] $ & $\alpha$  \\
   PuP    &   3.2                        &    1.76           &           1.60                                   &      -1.52                                &       0.08               &       0.75        &  0.82   \\
     PuAs    &  3.6                    &   1.60            &           1.67                                  &       -1.54                               &       0.13               &       0.64       &    0.86   \\
    PuSb  &   4.1                     &    1.40              &           1.57                                   &      -1.47                            &
 0.1                &        0.67       &   0.9 \\
 \hline  \hline 
   Antiferromagnetic \\
\hline \\
    &  U~[eV]  & $\lambda$~[a.u$^{-1}$]  &  $M_{\mathrm{s}}~[\mu_B] $ & $M_{\mathrm{orb}}~[\mu_B] $ & $M_{\mathrm{tot}}~[\mu_B] $ &  $M^{\mathrm{exp}}_{\mathrm{tot}}~[\mu_B] $ & $\alpha$  \\
   PuBi    &   4.1                       &     1.42           &           1.38                                  &      -1.29                             &     0.09               &  0.50            &  0.89   \\    
\hline \hline
\end{tabular}
\label{Tab:Pu-Comp-Mom}
\end{table}

\begin{thebibliography}{99}
\bibitem{HB}  J. Rossat-Mignod, G.~Lander and P.~Burlet, Hand-book on the Physics and Chcmistry of the Actinidcs, Vol.~1, eds. A.J.~Freeman and G.H. Lander~(North-Holland, Amsterdam, 1984) and references therein.
\bibitem{Heavy-fermions} G.R.~Stewart, Rev.~Modern Phys.~{\bf 56}, 755 (1984).
\bibitem{LDA-fails-I} T.~Kraft, P.M.~Oppeneer, V.N.~Antonov and H.~Eschrig, \prb~{\bf 52}, 3561 (1995). 
\bibitem{LDA-fails-II} P.M.~Oppeneer, M.S.S.~Brooks, V.N.~Antonov, T.~Kraft and H.~Eschrig, \prb~{\bf 53}, R10437 (1996).
\bibitem{Brooks} M.S.S.~Brooks, Physica {\bf 130B}, 6 (1985).
\bibitem{LAZ} A.I.~Liechtenstein, V.I.~Anisimov and J.~Zaanen, \prb~{\bf 52}, R5467 (1995).
\bibitem{Solovyev-I} I.V.~Solovyev, A.I.~Liechtenstein and K.~Terakura, \prl~{\bf 80}, 5758 (1998).
\bibitem{Anisimov} V.I.~Anisimov, I.V.~Solovyev, M.A.~Korotin, M.T.~Czy\.{z}yk and G.A.~Sawatzky \prb~{\bf 48}, 16929 (1993).
\bibitem{Petukhov} A.G.~Petukhov, I.I.~Mazin, L.~Chioncel and A.I.~Lichtenstein, \prb~{\bf 67}, 153106 (2003).
\bibitem{APW+lo-I} E.~Sj\"ostedt, L.~Nordstr\"om and D.~Singh,  Solid State Commun. {\bf 114}, 15 (2000).
\bibitem{APW+lo-II} D.~Singh, \prb~{\bf 43}, 6388 (1991)
\bibitem{Elk} Elk. An all-electron full-potential linearised augmented-plane wave plus local orbitals (FP-(L)APW+lo) code, available for free at \url{http://elk.sourceforge.net}.
\bibitem{Laskowski} R.~Laskowski, G.~K.~H.~Madsen, P.~Blaha, and K.~Schwarz,
\prb~{\bf 69}, 140408(R) (2004).
\bibitem{Shick_UGe2} A.B.~Shick and W.E.~Pickett, \prl~{\bf 86}, 300 (2001).
\bibitem{Shick} A.B.~Shick, V.~Drchal and L.~Havela, \epl~{\bf 69}, 588 (2005).
\bibitem{Shishidou} T.~Shishidou, T.~Oguchi and T.~Jo, \prb~{\bf 59}, 6813 (1999).
\bibitem{Norman} M.R.~Norman  \prb~{\bf 52}, 1421 (1995).
\bibitem{Judd} B.R.~Judd, Operator Techniques in Atomic Spectroscopy'', McGraw-Hill (1963). 
\bibitem{Brooks_Fks} M.S.~Brooks J. Phys. Cond. Matt.~{\bf 13}, L469 (2001).
\bibitem{Solovyev-II} I.V.~Solovyev, \prl {\bf 95}, 267205 (2005).
\bibitem{Czyzyk} M.T.~Czy\.{z}yk and G.A.~Sawatzky, \prb~{\bf 49}, 14211 (1994).
\bibitem{Slater} J.C.~Slater \pr {\bf 34}, 1293 (1929).
\bibitem{Racah-I} G.~Racah \pr {\bf 61}, 186 (1942).
\bibitem{Racah-II} G.~Racah \pr {\bf 62}, 438 (1942).
\bibitem{Racah-III} G.~Racah \pr {\bf 63}, 367 (1943).
\bibitem{Racah-IV} G.~Racah \pr {\bf 76}, 1352 (1949).
\bibitem{CS} E.U.~Condon and G.H.~Shortley, The Theory of Atomic Spectra, The University Press, Cambridge (1935), pp 174 of 1963 edition.
\bibitem{vdLaan1} G.~van~der~Laan and B.T.~Thoole, J. Phys. Cond. Matt.~{\bf 7}, 9947 (1995). 
\bibitem{Fano} U.~Fano, Rev. Mod. Phys.~{\bf 29}, 74 (1957).
\bibitem{Biedenharn} L.C.~Biedenharn, Annals of Physics~{\bf 4}, 104 (1958).
\bibitem{parity} The time reversal parity of a given tensor moment component $kpr$ is given by $(-1)^{k+p}$.
\bibitem{US-111-I}  D.L.~Tillwick and P.~de~V.~du~Plessis, J. Magn. Magn. Mater.~{\bf 3}, 319 (1976).
\bibitem{US-111-II} G.H. Lander, M.S.S. Brooks, B. Lebech, P.J. Brown, O. Vogt, and K. Mattenberger, J. Appl. Phys.~{\bf 69}, 4803 (1991).
\bibitem{US-It-I} J. Schoenes, B. Frick, and O. Vogt, \prb~{\bf 30}, 6578 (1984).
\bibitem{US-It-II} H. Rudigier, H.R. Ott, and O. Vogt, \prb~{\bf 32}, 4584 (1985).
\bibitem{US-It-III} B. Reihl, J. Less-Common Met.~{\bf 128}, 331 (1987).
\bibitem{US-Mom-5f} F.A. Wedgwood, J. Phys. C~{\bf 5}, 2427 (1972).
\bibitem{US-Mom}  G.~Busch, O.~Vogt, A.~Delpalme and G.H.~Lander, J. Phys. C~{\bf 12}, 1391 (1979).
 \bibitem{Cricchio} F.~Cricchio, F.~Bultmark and L.~Nordstr\"om, \prb~{\bf 78}, 100404(R)  (2008).
\bibitem{Granas} O.~Gr\aa n\"as, F.~Bultmark, F.~Cricchio and L.~Nordstr\"om, preprint.
\bibitem{Shick_PuCoGa5} P.M.~Oppeneer, A.B.~Schick, J. Rusz, S.~Lebegue and O.~Eriksson, J. of All. and Comp.~{\bf 444-445}, 109 (2007)
\bibitem{Brewer} L.~Brewer, J. Opt. Soc. Am.~\textbf{61}, 1101 (1971).
\bibitem{Lang} J.K. Lang, Y. Baer, and P.A. Cox, J. Phys F~\textbf{11}, 121 (1981).
\bibitem{vdLaan2} G.~van der Laan {\em et al.}, \prl~{\bf 93}, 097401 (2004).
\bibitem{Moore1} K.T.~Moore, G.~van~der~Laan, M.A.~Wall, A.J. Schwartz and R.~G.~Haire, \prb~{\bf 76}, 073105 (2007).
\bibitem{Moore2} K.T.~Moore and G.~van~der~Laan, \rmp~{\bf 81}, 235 (2009).

\end{thebibliography}
\end{document}